\documentclass[runningheads]{llncs}
\usepackage{graphicx}
\usepackage{enumerate}
\usepackage{pifont}
\usepackage{booktabs}
\usepackage{amssymb}
\usepackage{float}
\usepackage{booktabs}
\usepackage{multirow}
\usepackage[colorlinks=true,
            linkcolor=blue,
            filecolor=blue,      
            urlcolor=blue,       
            citecolor=blue
            ]{hyperref}
\usepackage{color, soul, framed}
\usepackage{makecell}
\usepackage[misc]{ifsym} 
\begin{document}
\title{DS$^3$-Net: Difficulty-perceived Common-to-T1ce Semi-Supervised Multimodal MRI Synthesis Network}
\titlerunning{DS$^3$-Net: Common-to-T1ce Modality Synthesis}
%
\author{Ziqi Huang\inst{1} \and 
Li Lin\inst{1,2}\and
Pujin Cheng\inst{1} \and
Kai Pan\inst{1} \and
Xiaoying Tang$^{1(\textrm{\Letter)}}$}
\authorrunning{Z. Huang et al.}
\institute{Department of Electrical and Electronic Engineering, Southern University of Science and Technology, Shenzhen, China\\ 
\email{tangxy@sustech.edu.cn} \and
Department of Electrical and Electronic Engineering, The University of Hong Kong, Hong Kong SAR, China}

%
\maketitle              
\begin{abstract}
Contrast-enhanced T1 (T1ce) is one of the most essential magnetic resonance imaging (MRI) modalities for diagnosing and analyzing brain tumors, especially gliomas. In clinical practice, common MRI modalities such as T1, T2, and fluid attenuation inversion recovery are relatively easy to access while T1ce is more challenging considering the additional cost and potential risk of allergies to the contrast agent. Therefore, it is of great clinical necessity to develop a method to synthesize T1ce from other common modalities. Current paired image translation methods typically have the issue of requiring a large amount of paired data and do not focus on specific regions of interest, e.g., the tumor region, in the synthesization process. To address these issues, we propose a Difficulty-perceived common-to-T1ce Semi-Supervised multimodal MRI Synthesis network (DS$^3$-Net), involving both paired and unpaired data together with dual-level knowledge distillation. DS$^3$-Net predicts a difficulty map to progressively promote the synthesis task. Specifically, a pixelwise constraint and a patchwise contrastive constraint are guided by the predicted difficulty map. Through extensive experiments on the publicly-available BraTS2020 dataset, DS$^3$-Net outperforms its supervised counterpart in each respect. Furthermore, with only 5\% paired data, the proposed DS$^3$-Net achieves competitive performance with state-of-the-art image translation methods utilizing 100\% paired data, delivering an average SSIM of 0.8947 and an average PSNR of 23.60. The source code is available at \url{https://github.com/Huangziqi777/DS-3_Net}.

\end{abstract}
\section{Introduction}
Magnetic resonance imaging (MRI) is a non-radioactive clinical imaging method that is widely used in gliomas diagnosis \cite{liu2015multimodalMRI,icsin2016review}. In clinical practice, there are some relatively easy-to-obtain MRI modalities such as T1, T2 and fluid attenuation inversion recovery (FLAIR), since they can be simultaneously acquired in a single routine MRI scanning process by adjusting specific parameters. Contrarily, there are some difficult-to-obtain MRI modalities such as contrast-enhanced T1 (T1ce), since they require external intervention (e.g., injection of a contrast agent) exerted to the patient of interest. Compared with common MRI modalities, T1ce can clearly distinguish the parenchymal part of brain tumor, which is of great guiding significance for the diagnosis of tumor-related diseases. However, the contrast agent is usually radioactive and some patients are allergic to it, and there is also an increase in the scanning cost. As such, paired MRI data of common and T1ce modalities are very scarce but indispensable in clinical practice, especially in tumor diagnosis. In such context, there has been great interest in research topics related to missing modality \cite{shen2019brain,zhou2020brain,zhou2021latent,shenfastmri}.

Deep learning (DL) has been widely employed in image synthesis tasks. For natural images, Pix2pix \cite{pix2pix} achieves good performance on paired image translation, while CycleGAN \cite{zhu2017unpaired} and CUTGAN \cite{park2020contrastive} perform well on unpaired image translation. For medical images, pGAN \cite{pgan} incorporates a content-preserving perceptual loss into Pix2pix and applies it to translation across 2D brain MRI. MedGAN \cite{armanious2020medgan} takes into account image style transformation while maintaining content consistency, which performs well on the PET-versus-CT conversion task. MM-GAN \cite{mmgan} randomly drops out several input modalities and makes use of selective discriminators for the loss calculation, synthesizing high-quality images under a missing-modality setting. However, these existing DL methods rely heavily on a large number of annotated or paired images in the training phase for supervision. Recently, semi-supervised learning which can fully exploit unlabeled data gains increasing interest, especially in the  medical image analysis realm which severely suffers from scarce data \cite{ibrahim2020semi,zhou2020deep,chen2021semi}.

Synthesizing MRI containing brain tumor is a very challenging task. Conventional image synthesis frameworks \cite{li2019diamondgan,yurt2021mustgan} for synthesizing glioma-contained MRI, especially for synthesizing T1ce and FLAIR, are usually unsatisfactory, mainly because of the following reasons: 1) Huge domain gaps exist across different glioma-contained MRI modalities, manifesting in the intensity distribution. 2) A large amount of paired data are required for well training a pixel-level image reconstruction network, which contradicts the real-life clinical situation of scarce paired data. 3) The image pattern of the glioma region varies significantly from those of other brain structures, which makes it difficult to be synthesized even if other brain structures can be well reconstructed.

In this work, we propose a novel framework for synthesizing the difficult-to-obtain MRI modality T1ce, employing semi-supervised learning with limited paired data and a larger number of unpaired data. Our main contributions are four-fold: 1) To the best of our knowledge, this is the first semi-supervised framework applied to multimodal MRI synthesis for gliomas. 2) In light of the teacher-student network, we make full use of unpaired multimodal MRI data through maintaining consistency in spaces of both high and low dimensions. 3) We innovatively estimate a difficulty-perceived map and adopt it to dynamically weigh both pixelwise and patchwise constraints according to the difficulty of model learning, and thus the model can assign the difficult-to-learn parts (such as the glioma region) more attention. 4) Extensive comparison experiments are conducted, both quantitatively and qualitatively.





\begin{figure}[t]
\centering
\centerline{\includegraphics[width=12cm]{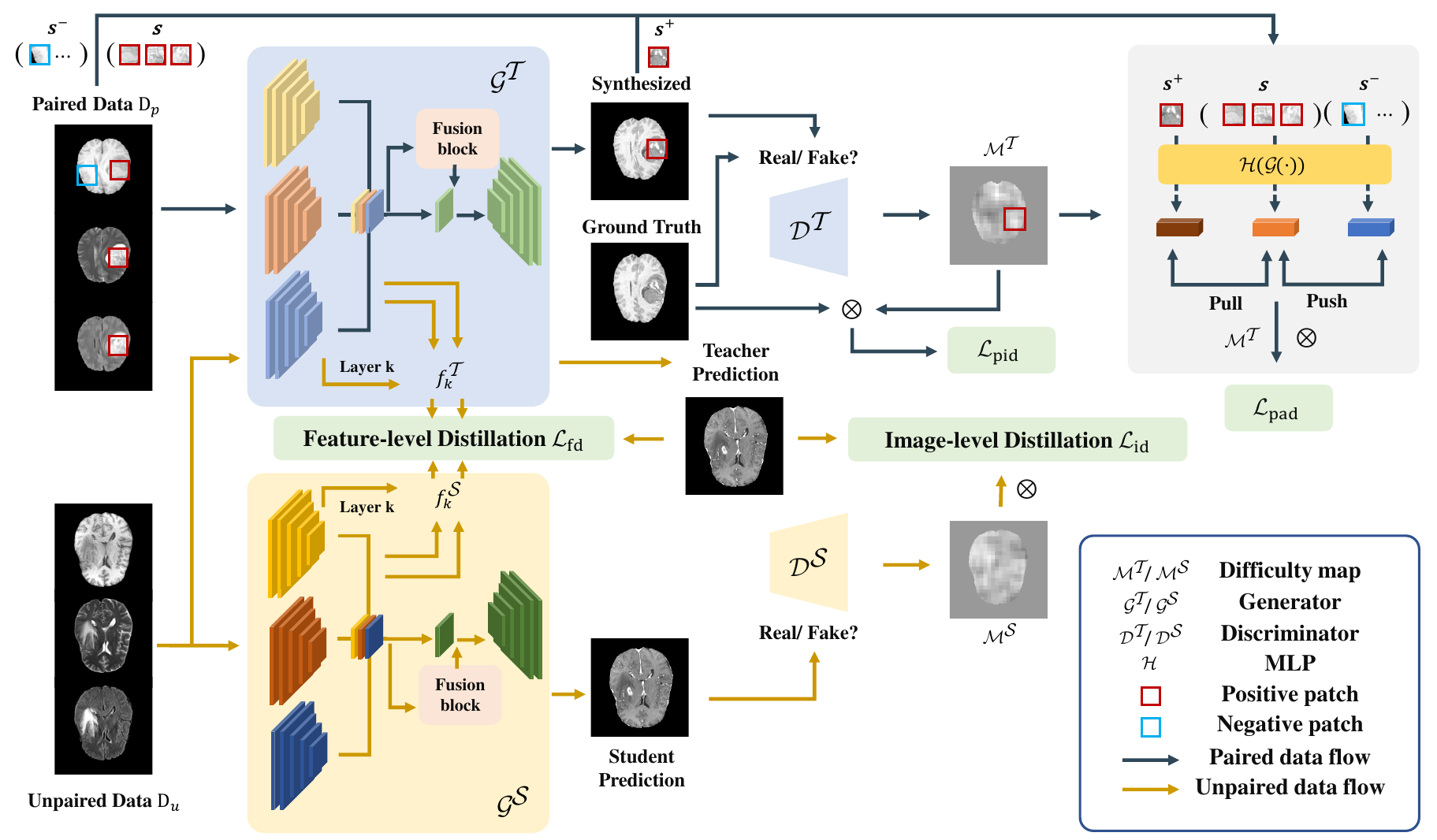}}
\vspace{-0.05cm}
\caption{Flowchart of our proposed DS$^3$-Net. For simplicity, the GAN losses for the teacher network and the student network, and the patchwise difficulty-perceived contrastive loss which is similar between the teacher network and the student network, are not included in this figure.} 
\label{flowchart} 
\end{figure}
\section{Methodology}
Let $D_p = \left \{\left ( x_i^\mathcal{T},y_i^\mathcal{T} \right )   \right \}_{i=1}^N$ denote the paired image dataset including the common MRI modalities $x_i^\mathcal{T} = \left \{\left ( _{1}x_i^\mathcal{T},_2x_i^\mathcal{T},..._nx_i^\mathcal{T}\right )   \right \}_{n=1}^C$, where $C$ is set as three including T1, T2 and FLAIR and $y_i$ is the target modality which is T1ce in our case. Let $D_u = \left \{\left ( x_i^\mathcal{S}\right )   \right \}_{i=1}^M$ denote the unpaired image dataset, with no corresponding T1ce. Note that the sample size of the unpaired datset $M$ is much larger than that of the paired datset $N$, with no patient-level overlap.

Our proposed Difficulty-perceived common-to-T1ce Semi-Supervised multimodal MRI Synthesis network (DS$^3$-Net) is demonstrated in Fig. \ref{flowchart}, including a teacher model for learning expertise knowledge from paired data and a student model for imitating the teacher's decision. The student model has the same architecture as the teacher model and the two models are encouraged to output consistent predictions. 
\subsection{Difficulty-perceived Attention Map}
One of the main challenges in conducting image synthesis to brain MRI containing glioma is that different regions have different synthesizing difficulties. For instance, normal brain tissues have relatively more straightforward intensity mapping relationships, while brain tumor regions often suffer from diverse patterns between common MRI and T1ce. Introducing an attention map to assess the synthesizing difficulty for each region is a feasible solution \cite{difficultymap}. Therefore, we intuitively utilize a predicted map $\mathcal{M}$ from the discriminator to dynamically guide the synthesis process, as the well-generated regions can easily fool the discriminator while the poorly-generated regions can not. In our case, the discriminator is a 70$\times$70 PatchGAN \cite{li2016precomputed}, leading to patch-level difficulty scores. Specifically, the difficulty map $\mathcal{M}^{\mathcal{T}}$ from the teacher model is defined as 
\begin{equation}
\mathcal{M}^\mathcal{T}=\left\{
		\begin{array}{lr}
		\left\|  1- \mathcal{D}_\mathcal{T}  \left ( \mathcal{G}_\mathcal{T}\left ( x_i^\mathcal{T} \right )   \right )    \right \|,&      x_i^\mathcal{T} > 0 \\ 
		0.2,                             &     x_i^\mathcal{T} = 0.
	    \end{array}
	\right.
\end{equation}
$\mathcal{D}_\mathcal{T}$ and $\mathcal{G}_\mathcal{T}$ respectively refer to the teacher network's discriminator and generator. For background pixels without any brain tissue, we set a small constant 0.2 to ensure that their synthesis results are normal. 

\subsection{Difficulty-perceived Pixelwise and Patchwise Constraints}
To fully utilize the multimodal information, we use three encoders each of which has nine ResNet blocks. We also employ fusion block modified from the SE module \cite{SEmodule} by replacing sigmoid with softmax to better fuse the three common modalities before being inputted to the decoder. 

L1 loss is adopted to provide a pixelwise constraint between the real T1ce $y_i^\mathcal{T}$ and the synthesized T1ce $\mathcal{G}_{\mathcal{T}}(x_i^\mathcal{T})$. We use the difficulty map $\mathcal{M}^{\mathcal{T}}$ to guide the L1 loss by assigning each pixel a difficulty score,  allowing the poorly synthesized regions to gain more attention. Specifically, we introduce a pixelwise difficulty-perceived L1 loss $\mathcal{L}_{pid}^\mathcal{T}$ for the teacher network
\begin{equation}
\label{lpid}
\mathcal{L}_{pid}^\mathcal{T} = \mathcal{M}^\mathcal{T} \circ \left \| y_i^\mathcal{T}-\mathcal{G}_{\mathcal{T}}(x_i^\mathcal{T}) \right \|_1.
\end{equation}

For the purpose of maintaining the same semantic information at the same location in a single sample, we use patchwise contrastive learning to constrain the parameters of different hidden layers. Specifically, we construct a positive pair $ \left ( z_{l}^s , z_{l}^{s+} \right ) $ and a negative pair $\left ( z_{l}^s , z_{l}^{s-} \right ) $, where $z_{l}^s=\mathcal{H}_{(l)}( \mathcal{G}_{\mathcal{T}(l)}(x_i^s))$ refers to features extracted from patch $s$ of the input $x_i$ at the $l^{th}$ layer of the teacher network's generator $\mathcal{G}_\mathcal{T}$. Next, we map the selected features into a high-dimension space by an additional MLP $\mathcal{H}_{(l)}$. Similarly, $z_{l}^{s+}=\mathcal{H}_{(l)}( \mathcal{G}_{\mathcal{T}(l)}(\hat{y}_i^s))$ refers to features from the synthesized T1ce $\hat{y}_i$ at patch $s$, regarded as the positive sample, and $z_{l}^{s-}$ represents features from patches different from $s$. We then pull the positive pair together and push the negative pair away via the InfoNCE loss \cite{van2018representation}. The difficulty map $\mathcal{M}^{\mathcal{T}}$ is down-sampled to match the resolution of each hidden layer $l^{th}$, defined as $m_{l}^\mathcal{T}$, and is multiplied with each patch's features to re-distribute the patchwise importance. To be specific, the patchwise difficulty-perceived contrastive loss $\mathcal{L}_{pad}^\mathcal{T}$ is calculated as 
\begin{equation}
\label{lpad}
\mathcal{L}_{pad}^\mathcal{T}=-\sum_{l\in L} m_{l}^\mathcal{T}  \circ \log \left[\frac{\exp \left(z_{l}^s \cdot z_{l}^{s+} / \tau\right)}{\left.\exp \left(z_{l}^s \cdot z_{l}^{s+} / \tau\right)+\sum \exp \left(z_{l}^s \cdot z_{l}^{s-} / \tau\right)\right)}\right],
\end{equation}where $L=\left \{0,4,8,12,16\right \} $ are the selected hidden layers and the temperature $\tau$ for the InfoNCE loss is empirically set as 0.07. 

Furthermore, an LSGAN loss \cite{lsgan} is adopted, encouraging the synthesized distribution to be close to the real one. In summary, the total loss function $\mathcal{L}^\mathcal{T}$ of the teacher network is 
\begin{equation}
\label{lsup}
\mathcal{L}^\mathcal{T}= \lambda_{pid}\mathcal{L}_{pid}^\mathcal{T}+\lambda_{pad}\mathcal{L}_{pad}^\mathcal{T}+\lambda_{GAN}\mathcal{L}_{GAN}^\mathcal{T},
\end{equation}where $\lambda_{pid}$, $\lambda_{pad}$ and $\lambda_{GAN}$ are empirically set as 100, 1, 1.

\subsection{Difficulty-perceived Dual-level Distillation}
We feed the unpaired dataset $D_u$ into the student network $\mathcal{N}_\mathcal{S}$ which is initialized with pre-trained weights from the teacher network. For semi-supervised learning, knowledge distillation \cite{hinton2015distilling,zhou2020deep} between the teacher and student models is one of the most common procedures to expand the dataset and perform parameter regularization. In our DS$^3$-Net pipeline, both feature-level and image-level distillations are 
performed to ensure the consistency of the two models in both high-dimension and low-dimension spaces. 

To be specific, we first input $x_i^\mathcal{S} \in D_u$ into both $\mathcal{N}_\mathcal{S}$ and $\mathcal{N}_\mathcal{T}$ to obtain a coarse prediction $\mathcal{G}_{\mathcal{S}}(x_i^\mathcal{S})$ and a fine prediction $\mathcal{G}_{\mathcal{T}}(x_i^\mathcal{S})$. Given that the paired ground truth for $x_i^s$ is not available, we treat the fine decision $\hat{q}_i^{\mathcal{T}}$ as the pseudo ground truth for the student network, and then conduct image-level distillation through an L1 loss function. 

We expect the student model to imitate the teacher model in terms of representation at the high-dimension space to better handle the unpaired samples. Therefore, when one sample goes through the two networks, we extract the $k^{th}$ layer's features $f_k^\mathcal{T}$ and $f_k^\mathcal{S}$, and then perform feature-level distillation between the two networks by minimizing the feature discrepancy.

Apparently, the student discriminator $\mathcal{D}_\mathcal{S}$ can also produce a difficulty map $\mathcal{M}^\mathcal{S}$ for each unpaired sample and we employ the same difficulty guided strategy towards the distillation process. We thus have the image-level distillation loss
\vspace{+0.02cm}
\begin{equation}
\label{lid}
\mathcal{L}_{id}^\mathcal{S}= \mathcal{M}^\mathcal{S} \circ \left \| \mathcal{G}_{\mathcal{T}}(x_i^\mathcal{S})-\mathcal{G}_{\mathcal{S}}(x_i^\mathcal{S}) \right \|_1,
\end{equation} and the feature-level distillation loss
\begin{equation}
\label{lfd}
\mathcal{L}_{fd}^\mathcal{S}= \frac{1}{K}\sum_{k \in L_{distill}} m_{k}^\mathcal{S} \circ  \left \| f_k^\mathcal{T}-f_k^\mathcal{S} \right \|_1, 
\end{equation}where $L_{distill}=\left \{4,8,12,16,21 \right \}$, $K$ denotes the number of the selected layers, and $m_{k}^\mathcal{S}$ represents the difficulty map down-sampled to the same size as that of the feature map of the $k^{th}$ layer. 

Similar to the teacher model, the student model also has an LSGAN loss $\mathcal{L}_{GAN}^\mathcal{S}$ and a patchwise difficulty-perceived contrastive loss $\mathcal{L}_{pad}^\mathcal{S}$. Collectively, the loss function $\mathcal{L}^\mathcal{S}$ for the student network is defined as
\begin{equation}
\label{lstu}
\mathcal{L}^\mathcal{S}= \lambda_{id}\mathcal{L}_{id}^\mathcal{S}+
\lambda_{fd}\mathcal{L}_{fd}^\mathcal{S}+\lambda_{pad}\mathcal{L}_{pad}^\mathcal{S}+\lambda_{GAN}\mathcal{L}_{GAN}^\mathcal{S}.
\end{equation}Here we empirically set $\lambda_{id}$, $\lambda_{fd}$, $\lambda_{pad}$ and $\lambda_{GAN}$ as 100, 1, 1 and 1.

\vspace{-0.15cm}
\section{Experiments and Results}
\subsection{Dataset}
BraTS2020 \cite{brats2020} (\url{https://www.med.upenn.edu/cbica/brats2020/}) is a large publicly-available dataset, consisting in total 369 paired MRI data of  T1, T2, FLAIR, and T1ce from gliomas patients. We divide each 3D image into 2D slices at the axial view. Each 2D slice is center-cropped to 224 $\times$ 224. We divide the entire dataset into training, validation, and test sets according to a ratio of 7:1:2 at the patient level, and use slices that contain tumor during the training phase. Since the tumor core (TC) containing the enhancing tumor (ET) is the key region to distinguish T1ce from other modalities while it usually constitutes only a small portion of the whole tumor, we adjust the proportion to be 1:1 for the numbers of slices with and without TC.


\vspace{-0.06cm}
\subsection{Training Strategy}
\vspace{-0.06cm}
At the first training stage, the paired dataset are utilized to pre-activate the teacher network, and we take the weights of the 5$^{th}$ epoch before convergence as the initialization weights for the next stage of training to prevent overfitting. Specifically, each difficulty map score is set to be the constant 1 in the first stage which ensures a stable output from the initial network. In the second stage, we start with the difficulty maps delivered by the discriminators and train the two networks together by optimizing the final loss function $\mathcal{L}=\mathcal{L}^\mathcal{T}+(1-\frac{t}{T})\mathcal{L}^\mathcal{S}$, where $t \in [0,T)$ represents the current epoch counting and $T$ is the total number of the training epochs. All methods are implemented with Pytorch using NVIDIA RTX 3090 GPUs. AdamW \cite{AdamW} is chosen to be the optimizer and the batch size is set as 6 with equivalent numbers of paired and unpaired samples. The learning rates in the first stage for the generator $\mathcal{G}^{\mathcal{T}}$, the MLP $\left \{ \mathcal{H}_{(l)} | l\in \mathcal{L} \right \}$ and the discriminator $\mathcal{D}^{\mathcal{T}}$ are respectively set as 0.0006, 0.0006 and 0.0003, and these values are attenuated to a fifth of the original numbers in the second training phase. For the student network training, we have the same learning rate setting as that for the initial teacher network and train for a total of 100 epochs. All of the learning rate settings follow the linear decay strategy.

\begin{table}
\renewcommand\arraystretch{1.05}
\centering
\caption{Quantitive evaluations for DS$^3$-Net with different percentages of paired data. Bold metric represents the best performance.}
\label{table1}
\begin{tabular}{c|c|ccc|cc} 
\bottomrule[1pt]
\multirow{2}{*}{\begin{tabular}[c]{@{}c@{}}Paired \\ percentage\end{tabular}} & \multirow{2}{*}{Method}  & \multicolumn{3}{c|}{Image quality}                & \multicolumn{2}{c}{Segmentation quality}  \\ 
\cline{3-7}
                                                                              &                          & SSIM$\uparrow$   & PSNR$\uparrow$  & MSE$\downarrow$   & ET Dice$\uparrow$     & TC Dice$\uparrow$          \\
\Xhline{1pt}
0\%                                                                                                                                                   & CUTGAN \cite{park2020contrastive}   &   0.7982               &17.63 &0.020  & 33.9\% & 56.1\% \\
\hline
\multirow{2}{*}{5\%}                                                            & DS$^3$-Net (paired only) & 0.8691                  & 21.97                   & \textbf{0.008}           & 39.5\%                     & 63.7\%                      \\
                                                                                & DS$^3$-Net               & \textbf{0.8947}         & \textbf{23.60}          & 0.009                    & \textbf{43.3\%}            & \textbf{66.0\%}             \\ 
\hline
\multirow{2}{*}{10\%}                                                           & DS$^3$-Net (paired only) & 0.8763                  & 23.01                   & \textbf{0.007}           & 43.6\%                     & 65.3\%                      \\
                                                                                & DS$^3$-Net               & \textbf{0.8943}         & \textbf{23.90}          & 0.008                    & \textbf{43.6\%}            & \textbf{66.9\%}             \\ 
\hline
\multirow{2}{*}{50\%}                                                           & DS$^3$-Net (paired only) & \textbf{0.8895}         & 23.73                   & 0.006                    & 44.1\%                     & 67.9\%                      \\
                                                                                & DS$^3$-Net               & 0.8893                  & \textbf{23.79}          & \textbf{0.005}           & \textbf{44.9\%}            & \textbf{68.1\%}             \\ 
\hline
\multirow{4}{*}{100\%}                                                          
                                                                              & pGAN \cite{pgan}                     & 0.8846               & 23.56                & 0.010                 & 36.7\%               & 61.8\%                \\
                                                                              & MedGAN \cite{armanious2020medgan}                  & 0.8900               & 24.34                & 0.011                 & 37.8\%               & 58.3\%               \\
                                                                              & Pix2pix \cite{pix2pix}                & \textbf{0.9007}      & \textbf{25.02}       & \textbf{0.004}        & 41.7\%               & 65.0\%                \\
\multicolumn{1}{l|}{}                                                          
                                                                              & DS$^3$-Net               & 0.8959               & 24.26                & 0.005                 & \textbf{47.8\%}      & \textbf{67.2\%}       \\
\bottomrule[1pt]
\end{tabular}
\end{table}

\vspace{-0.08cm}

\subsection{Results}
\vspace{-0.06cm}
In the training phase, we consider a few percentages (e.g., 5\%, 10\% and 50\%) of the full training set as our paired dataset and the rest as the unpaired dataset. To better evaluate our method, we compare DS$^3$-Net with several state-of-the-art (SOTA) methods including two for natural images (Pix2pix \cite{pix2pix} and CUTGAN \cite{park2020contrastive}) and two for medical images (pGAN \cite{pgan} and MedGAN \cite{armanious2020medgan}). 

The structure similarity index measure (SSIM), peak signal-to-noise ratio \\(PSNR) \cite{ssim} and mean squared error (MSE) are employed to evaluate the effectiveness of DS$^3$-Net.  Besides, because T1ce plays a crucial role in identifying ET and TC \cite{436brain1,436brain2,436brain3}, we perform tumor segmentation using the synthesized T1ce together with the three common modalities and use the Dice scores \cite{dice2015metrics} of ET and TC to assess the quality of the synthesized T1ce. The entire test set is divided into five folds, four of which are used for training the segmentation network and one for testing. UNet \cite{ronneberger2015u} is employed as our segmentation network. During the testing phase, we use the student network to deliver the prediction. All comparison results are listed in Table \ref{table1}.

\vspace{-0.05mm}
\begin{table}[b]
\centering
\caption{Ablation studies for each component of DS$^3$-Net. }
\begin{tabular}{cccc}
\toprule
\textbf{Component}                & \textbf{SSIM$\uparrow$}   & \textbf{PSNR$\uparrow$}  & \textbf{MSE$\downarrow$}   \\
\midrule
DS$^3$-Net w.o. $\mathcal{M}_\mathcal{T}/\mathcal{M}_\mathcal{S}$ & 0.8884 & 21.38 & 0.012 \\
DS$^3$-Net w.o. $\mathcal{L}_{fd}$           & 0.8883 & 21.33 & 0.012 \\
DS$^3$-Net w.o. $\mathcal{L}_{id}$            & 0.7424 & 13.43 & 0.063 \\
DS$^3$-Net teacher network $\mathcal{N}_{T}$       & 0.8715 & 20.67 &   0.013    \\
DS$^3$-Net student network $\mathcal{N}_{S}$                       & \textbf{0.8947} & \textbf{23.60} & \textbf{0.009}    \\
\bottomrule
\end{tabular}
\label{ablation1}
\end{table}

\begin{figure}[t]
\centering
\centerline{\includegraphics[width=12cm]{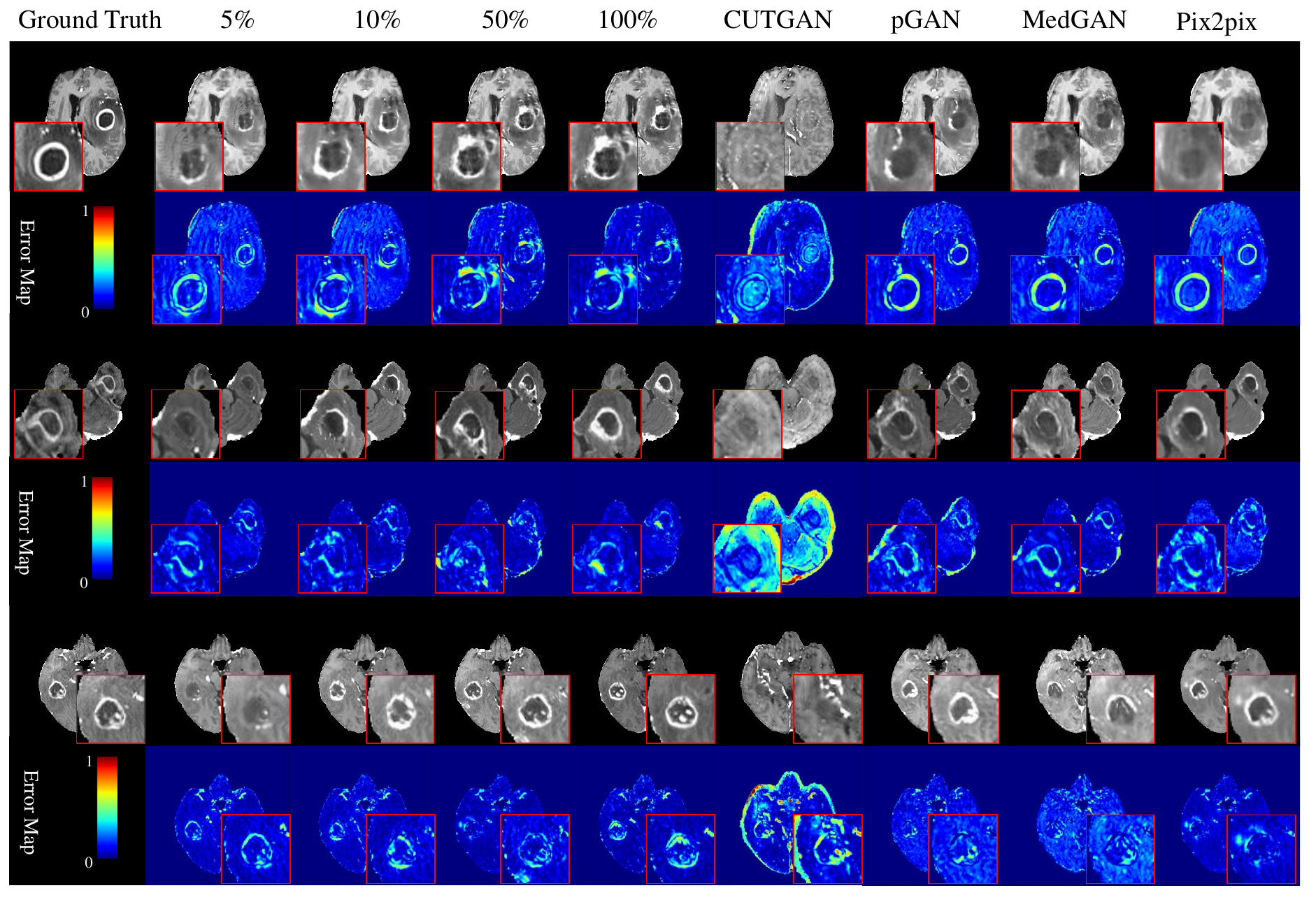}}
\vspace{-0.3cm}
\caption{Visualization of synthesized T1ce and the corresponding error maps from different methods. Each percentage represents DS$^3$-Net trained with different paired percentage.} 
\label{visual2} 
\end{figure}

We observe that the semi-supervised training strategy can effectively improve the synthesized T1ce's quality and the segmentation performance of ET and TC compared to the vanilla supervised method (namely DS$^3$-Net (paired only)), as demonstrated in Table \ref{table1}. Apparently, DS$^3$-Net achieves competitive performance with full-data training methods, even with only 5\% paired data for training, especially in segmenting ET and TC. To qualitatively illustrate the performance, we display visualizations of the synthesized T1ce from different methods in Fig. \ref{visual2}. It can be easily observed that DS$^3$-Net performs well in synthesizing tumor regions and achieves competitive performance with other SOTA methods trained with 100\% paired data. Error maps are also presented in Fig. \ref{visual2}, revealing that the erroneous regions mainly focus around the tumor, but our proposed DS$^3$-Net can effectively alleviate this dilemma to some extent (also see Fig. A1 of the appendix). Additionally, by zooming the tumor regions for visualization, DS$^3$-Net is found to deliver T1ce with a clearer boundary for ET compared with other SOTA methods.

Ablation studies are conducted with 5\% paired data training, to assess each component's importance in the proposed DS$^3$-Net. As shown in Table \ref{ablation1}, removing any of the proposed components deteriorates the synthesization performance in terms of all evaluation metrics. In addition, it can be found that without $\mathcal{L}_{id}$, DS$^3$-Net suffers dramatic degradation, revealing that image-level distillation is an indispensable component in our framework. Since the student network inputs more data and receives reliable guidance from the teacher network, the image quality is effectively improved compared with that from the teacher network.
\vspace{-0.05cm}
\section{Conclusion}
\vspace{-0.05cm}
In this work, we propose a novel difficulty-perceived semi-supervised multimodal MRI synthesis pipeline to generate the difficult-to-obtain modality T1ce from three common easy-to-obtain MRI modalities including T1, T2, and FLAIR. Difficulty-perceived maps are adopted to guide the synthesization process of important regions, and dual-level distillation enables the model to train a well-performimg network with limited paired data. Extensive experiments are conducted on a large publicly-available dataset, and the effectiveness of  DS$^3$-Net is successfully identified.

\end{document}